# Delay based Duplicate Transmission Avoid (DDA) Coordination Scheme for Opportunistic routing

Ning Li, *Student Member IEEE*, Jose-Fernan Martinez-Ortega, Vicente Hernandez Diaz

*Abstract*-Since the packet is transmitted to a set of relaying nodes in opportunistic routing strategy, so the transmission delay and the duplication transmission are serious. For reducing the transmission delay and the duplicate transmission, in this paper, we propose the delay based duplication transmission avoid (DDA) coordination scheme for opportunistic routing. In this coordination scheme, the candidate relaying nodes are divided into different fully connected relaying networks, so the duplicate transmission is avoided. Moreover, we propose the relaying network recognition algorithm which can be used to judge whether the sub-network is fully connected or not. The properties of the relaying networks are investigated in detail in this paper. When the fully connected relaying networks are got, they will be the basic units in the next hop relaying network selection. In this paper, we prove that the packet delivery ratio of the high priority relaying nodes in the relaying network has greater effection on the relaying delay than that of the low priority relaying nodes. According to this conclusion, in DDA, the relaying networks which the packet delivery ratios of the high priority relaying nodes are high have higher priority than that of the low one. During the next hop relaying network selection, the transmission delay, the network utility, and the packet delivery ratio are taken into account. By these innovations, the DDA can improve the network performance greatly than that of ExOR and SOAR.

*Index Term*-Opportunistic routing, coordination scheme, transmission delay, duplicate transmission.

I. INTRODUCTION

In the past decades, the wireless sensor networks (WSNs) have been applied more and more widely, such as in the wildlife monitoring [1][2][3][4], the forest protection [5][6][7], the smart grid [8][9][10][11][12][13], the smart city [14][15][16], etc. The WSNs change our lifestyle in many areas. One of the critical issues of WSNs is the routing algorithm design, which guarantees reliable and efficient data transmission between the source node and the destination node. The routing algorithms of WSNs have been investigated for a long time and many excellent algorithms have been proposed to improve the network performance. These algorithms can be divided into two categories: the deterministic routing and the opportunistic routing [17]. In deterministic routing, the source node chooses one of its neighbors as the next hop relaying node based on optimal algorithms. The advantages of the deterministic routing algorithm are that it is simple and the duplicate transmission is slight. However, in the deterministic routing, the packet delivery ratio between the sender and the receiver is low and varies, which cause frequent packet retransmission between sender and receiver. For solving this issue, the authors in [17] propose the concept of opportunistic routing. In opportunistic routing, the sender sends the data packet to a set of neighbors to improve the packet delivery ratio. The opportunistic routing can improve the packet delivery ratio successfully; however, due to more than one neighbors receive the data packet from the sender, so during the data packet relaying, the transmission delay and the duplicate transmission are higher than that of the deterministic routing strategy.

The opportunistic routing can be divided into two stages: 1) the sender chooses the candidate relay nodes and prioritizes these nodes based on some performance metrics (such as, the distance to the destination node, the ETX, the residual energy, etc); in this stage, the node utility, denoted as *U*, is calculated based on the performance metrics; 2) the nodes in the candidate relaying set relay the data packet to the next hop relaying nodes based on the coordination schemes (such as the time-based coordination scheme [17][18], the contention-based coordination scheme [19][20][21], etc). In the previous researches, the candidate relay nodes selection and prioritization have been investigated in detail [22]. The second stage is important to the routing performance, since the transmission delay and the duplicate transmission are mainly caused by this stage. Because in this stage, when the candidate relay nodes receive the data packet, who is the first one to transmit the data packet to the next hop relaying nodes and how they notify the other relaying nodes that the data packet has been relayed to the next hop relaying nodes are decided. There are four coordination schemes for the opportunistic routing: contention-based coordination, time-based coordination, token-based coordination [23][24], and random coordination [25]. In this paper, we mainly focus on the time-based coordination scheme.

The principle of the time-based coordination scheme has been introduced in detail in [17], [18], and [22]. The main issue with the timer-based solution is that it is based on packet overhearing, thereby leading to high duplicate transmissions and transmission delay [22]. These latter occur when some candidates do not overhear the selected relay's reply. This is the case especially in sparse networks, where candidate relays are placed further apart. In order to mitigate this problem, one possible solution consists in removing some nodes from the candidate relay set so that only fully connected candidate relays are kept.

Ning Li, Jose-Fernan Martinez-Ortega, and Vecente Hernandez Diaz are with the Universidad Politenica de Madrid, Madrid, Spain.
The research leading to the presented results has been undertaken with in the SWARMs European project (Smart and Networking Underwater Robots in Cooperation Meshes), under Grant Agreement n. 662107-SWARMs-ECSEL-2014-1, partially supported by the ECSEL JU and the Spanish Ministry of Economy and Competitiveness (Ref: PCIN-2014-022-C02-02).
E-mail: {li.ning, jf.martinez, vicente.hernandez}@upm.es.

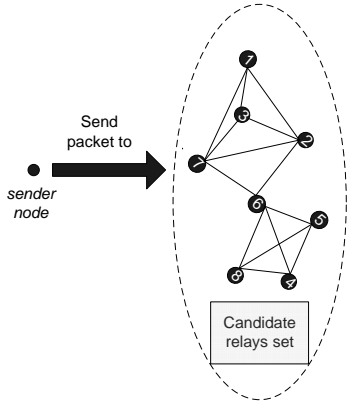

Fig. 1. The candidate relaying networks of opportunistic routing.

Unfortunately, in the previous researches, how to construct and judge the fully connected relaying network has not be investigated sufficiently. Moreover, as shown in Fig. 1, to the nodes in the candidate relaying set, more than one fully connected relaying networks can be constructed; the topologies and the nodes in these networks are different, such as the network (1,2,3,7) and network (4,5,8), etc. For the candidate relaying set shown in Fig. 1, many different relaying networks can be constructed; since the nodes and topologies in these relaying networks are different, so the properties (such as the relaying delay, the packet delivery ratio, etc) of these networks are different; for example, the packet delivery ratio and the relaying delay of networks (1,2,3,7) and network (4,5,8) are different. Therefore, how to evaluate the performance of these relaying networks and select the most appropriate relaying network for the opportunistic routing are also the main contents of this paper.

Moreover, for reducing the transmission delay, the node which the packet delivery ratio is high should have high relaying priority (this will be proved in Section IV). As the viewpoints proposed in [17] and [26], the packet deliver ratios of the nodes in the communication link from the source node to the destination node have different effection on the routing performance. For instance, the packet delivery ratio of the node at the end of the link have great effect on the energy consumption and transmission delay [26]; the ETX relates to all the packet delivery ratios of nodes in the communication link from the source node to the destination node [17]. In this paper, we will prove that the routing performance, such as the transmission delay, is also affected greatly by the first node's packet delivery ratio in the communication link. Therefore, in this paper, the effect of the packet delivery ratios of the candidate relaying nodes on the routing performance will be investigated in detail.

Motivated by these, we propose a new time-based coordination scheme, named the delay based duplicate transmission avoid (DDA) coordination scheme. In DDA, the nodes in the candidate relaying set are divided into different fully connected relaying networks. Since for the candidate relaying nodes, more than one relaying networks can be constructed and only one relaying networks can be chosen as the final relaying network, so the main objectives of this paper can be summarized as: 1) how to recognize the fully connected relaying networks that constructed by the candidate relaying nodes; and 2) how to chosen the most suitable relaying network from these networks. In DDA, the relaying network selection takes the packet delivery ratio between the sender and the relaying networks, the transmission delay, and the relaying priorities of nodes in the relaying networks into account to choose the most effective relaying network. By these, the transmission delay and duplicate transmission are reduced while the effective of the opportunistic routing is kept. The main contributions of this paper can be summarized as follows:

1. We define the relaying network for the candidate relaying set; the relaying networks are fully connected. Moreover, we also propose an algorithm to judge which nodes can construct a fully connected relaying networks and how many relaying networks can be constructed by the candidate relaying nodes; to the best of our knowledge, this is the first algorithm that introduce the relaying network into the relaying nodes selection and can be used to judge whether the network is fully connected or not;

2. We propose the calculation model of the one-hop average relaying delay for the opportunistic routing;

3. By investigating the relaying networks, we propose the inner-network properties and the in-network properties of the relaying networks, which can be used in the relaying network selection;

4. Based on the transmission delay of the relaying network and the packet delivery ratio between the sender and relaying network, we propose a relaying network selection algorithm to choose the most suitable relaying network for data packet transmission; in this algorithm, not only the transmission delay and the packet deliver ratio, but also the node utility and the relaying priority of node are taken into account.

The rest of this paper is organized as follows: Section II introduces the different coordination schemes of opportunistic routing; in Section III, the problems will be solved in this paper are stated, the network model is introduced, and the calculation model of network relaying delay and packet delivery ratio are proposed; Section IV investigates the properties of the relaying networks; in Section V, the relaying network based coordination scheme for the opportunistic routing is proposed; in Section VI, the performance of the new coordination scheme is compared with the traditional ones; the Section VII summaries the work in this paper.

## II. RELATED WORKS

The coordination scheme is used to find the appropriate candidate relaying nodes for packet transmission. In the past decades, many coordination schemes for the opportunistic routing algorithm have been proposed. These coordination schemes can select the best relaying node while incurring the smallest cost (in terms of the relaying delay, the duplicated transmission, etc) and can be classified into four main classes: contention-based coordination [19][20][21], time-based coordination [17][18], token-based coordination [23][24], and random coordination [25]. In the following of this section, we will introduce the algorithms relate to these four schemes briefly.

In [17] and [18], the time-based coordination schemes are used. In [17], the concept of opportunistic routing is proposed. The source node prioritizes and chooses the candidate relaying nodes based on the values of nodes' estimated transmission count (ETX) to the destination node. The node which the ETX is small will be set with high priority to relay the packet. When the source node sends the packet, the relaying nodes relay in the order in which they appear in the forwarding list, highest priority first. Low priority relaying nodes drop the packet when they receive the ACK from the high priority relaying nodes during the waiting time; otherwise, relaying the packet. Similar to [17], in [18], the coordination scheme is the same as that introduce in [17]; moreover, in this algorithm the waiting timer is set to 45*ms*. For reducing the transmission delay in time-based coordination schemes, some algorithms introduce the network coding into the routing algorithm, such as in [27], [28], [29], and [30]. The network coding improves the network throughput and reduces the overhead; however, the issue of deciding when and how often to generate coded packets is still not solved in these researches [22].

The contention-based coordination scheme is used in [19], [20], and [21]. In [19], the source node send RTF (Request to Forward) packet, the neighbors who receive this RTF packet will reply CTF (clear to forward) packet to the source node. These CTF packets' transmission is competitive with each other. The neighbor which the CTF packet is received by the source node will be the next hop relaying node. The forwarding scheme used in [20] is the similar contention-based scheme with that used in [19]. The coordination approach used in [21] is different with that shown in [19] and [20]. In [21], when the candidate relaying nodes receive the packet transmitted from the source node, they will content the same transmission channel to relay the packet to the next hop relaying nodes; the candidate relaying node which competes to the communication channel will transmit the packet to the next hop relaying nodes.

In the token-based coordination schemes, such as [23] and [24], since only the node which holds the token can transmit packets, so the duplicate transmission is reduced greatly. However, in the token-based coordination scheme, the control cost is pretty high. When the source node transmits the packet to the candidate relaying nodes, the relaying nodes receive and store the packet. The relaying node is allowed to relay the packet only it receives the tokens. The tokens include the acknowledgement information and are passed from high priority relaying node to low priority node. The candidate relaying nodes receive the tokens and can only transmit the unacknowledgement packet. Similar to the time-based coordination scheme, in the token-based coordination scheme, the candidate relaying nodes should also be fully connected.

For reducing the waiting delaying in the above coordination schemes, in [25], the authors propose the random selection coordination scheme. In this scheme, each candidate relaying node decides whether to continue forwarding the packet to the destination or not probabilistically, so the relaying delay that caused by the waiting timer is reduced greatly. However, in this scheme, since the candidate relaying nodes decide whether forwarding or not probabilistically, so the duplication transmission is serious.

### III. NETWORK MODEL

*A. Network model*

In this paper, two nodes can communicate with each other directly (without the help of the third node) if and only if there is a bi-directional communication link between these two nodes. The bi-directional communication link means that the transmission ranges of these two nodes are all larger than the distance between these two nodes. For instance, as shown in Fig. 2(a), node *s* and node 7 can communicate with each other directly when $\|s7\| \leq r_s$ and $\|s7\| \leq r_7$, where $\|s7\|$ is the Euclidean distance between node *s* and node 7, $r_s$ and $r_7$ are the transmission ranges of node *s* and node 7, respectively. The transmission range of node *s* is a circle which the centre is node *s* and the radius is $r_s$, denoted as $C(s, r_s)$. This can be found in Fig. 2(a).

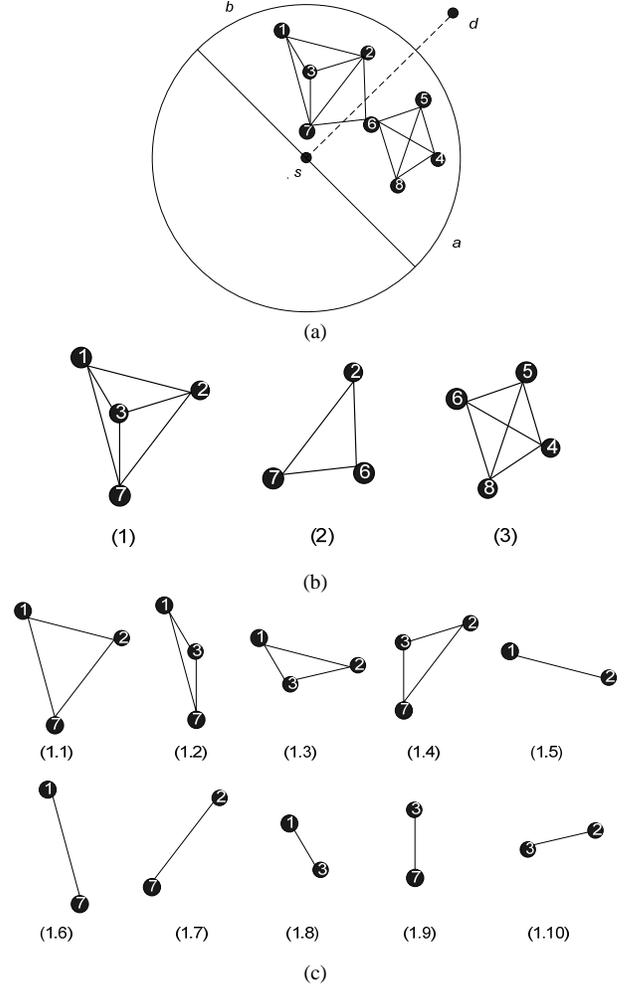

Fig. 2. The network model for opportunistic routing: (a) the network of the candidate relaying nodes; (b) the independent sub-networks of the original network; (c) the dependent sub-networks of Fig. 2(b.1)

As shown in the Fig. 2(a), in opportunistic routing, when the sender wants to send data packet, first, a set of neighbor nodes are chosen as the candidate relaying nodes based on the performance metrics (such as, ETX, distance, etc), and the sender relays the data packet to all the nodes in the candidate relaying set $\mathbb{R}$ (the candidate relaying set is the set of all the candidate relaying nodes). For instance, in Fig. 2(a), $\mathbb{R} = \{1,2,3,4,5,6,7,8\}$. The network that constructed by the nodes in $\mathbb{R}$ is denoted as $G(V_\mathbb{R}, E_\mathbb{R})$, where $V_\mathbb{R}$ represents the set of nodes in $\mathbb{R}$ and $E_\mathbb{R}$ represents the set of bi-directional communication links in the network. Second, the candidate relaying nodes relay the data packet to the next hop candidate relaying nodes with the same process as the sender. In this paper, we mainly concentrate on the second step.

In the second step, the candidate relaying nodes need to be filtered based on the requirements of the coordination schemes. For instance, in the time-based coordination scheme, the relaying nodes should be able to communicate directly with each other, i.e. the network constructed by these nodes should be fully connected. The fully connected network means that between any two nodes in this network there exists a bi-directional communication link; otherwise, the network is not fully connected. However, as shown in Fig. 2(a), the $G(V_\mathbb{R}, E_\mathbb{R})$ may not the fully connected network. For instance, node_3 and node_6 are not connected directly. The feasible approach is to keep the fully connected candidate relaying node set $\mathbb{R}^*$ and remove the un-fully connected nodes, where $\mathbb{R}^*$ is the subset of $\mathbb{R}$. For instance, the nodes in $\mathbb{R}^* = \{1,2,3,7\}$, which is shown in Fig. 2(b), are fully connected. To $\mathbb{R}$, there are many different subsets $\mathbb{R}^*$, which means that to $G(V_\mathbb{R}, E_\mathbb{R})$, there are many fully connected sub-networks $G(V_{\mathbb{R}^*}, E_{\mathbb{R}^*})$ can be constructed by the nodes in $\mathbb{R}$. For example, the networks are shown in Fig. 2(b) and Fig. 2(c) are all the fully connected sub-networks of Fig. 2(a). Since these fully connected networks are different, so for investigating the differences between these networks more clearly, some definitions are presented as follows.

In the fully connected networks, there must have bi-directional links between any two nodes, so we can simplify the expression of the fully connected network by only showing the nodes in this network; such as, for $G((2,6,7),(\overline{26},\overline{27},\overline{67}))$ which is shown in Fig. 2(b.2), we can simplify the expression as $G(2,6,7)$. Since in the time-based coordination scheme, the relaying networks that constituted by the candidate relaying nodes should be fully connected, so we define the relaying network as follows.

**Definition 1:** The fully connected sub-networks $G(V_{\mathbb{R}^*})$ of $G(V_\mathbb{R})$ are the relaying networks of candidate relaying set $\mathbb{R}$.

For instance, in Fig. 2(a), $G(2,6,7)$ is one of the relaying networks. Since there are more than one relaying networks and the nodes in these relaying networks are different, such as the relaying networks $G(2,6,7)$ and $G(1,2,3,7)$, so for distinguishing these networks, we define the network degree in Definition 2.

**Definition 2:** The degree of the relaying network is defined as the number of nodes in the relaying networks, denoted as $d_G$.

For instance, in Fig. 2(b), the network degree of Fig. 2(b.1) is 4. Notice the fact that in the relaying networks, the small degree relaying networks may be the sub-network of the large degree relaying networks (it is not always true); so we define the relevant and irrelevant for the relaying networks in Definition 3.

**Definition 3:** For any two relaying networks $G(V_{\mathbb{R}_1^*})$ and $G(V_{\mathbb{R}_2^*})$, in which $V_{\mathbb{R}_1^*} \notin V_{\mathbb{R}_2^*}$ and $V_{\mathbb{R}_2^*} \notin V_{\mathbb{R}_1^*}$, if $G(V_{\mathbb{R}_1^*} + V_{\mathbb{R}_2^*})$ is still the relaying network, then these two relaying networks are relevant; otherwise, these two relaying networks are irrelevant.

Based on Definition 3, we can give the Definition 4 as follows.

**Definition 4:** For the relaying network $G(V_{\mathbb{R}_i^*})$, if there exist relaying network $G(V_{\mathbb{R}_j^*})$ which relevant with $G(V_{\mathbb{R}_i^*})$, then $G(V_{\mathbb{R}_i^*})$ is called *s-network*; otherwise, $G(V_{\mathbb{R}_i^*})$ is called *o-network*.

For instance, the $G(1,2,3,7)$ shown in Fig. 2(b) is an *o*-network; the $G(1,2,3)$ shown in Fig. 2(c) is a *s*-network of $G(1,2,3,7)$. The *s*-network can be derived from the *o*-network. To each *o*-network, there are more than one *s*-networks can be derived from this *o*-network; the degree of these *s*-networks are smaller than that of the *o*-network. For instance, the relaying networks shown in Fig. 2(c) are all *s*-networks that derived from the *o*-network shown in Fig. 2(b.1). Moreover, since the network degree of Fig. 2(b.1) is 4, so the *s*-networks that derived from Fig. 2(b.1) will be 2-degree and 3-degree, respectively. Note that if the network degree is 1-degree, then the algorithm will be the same as the deterministic routing, so in this paper, we do not consider the 1-degree networks. The notations used in this paper are listed in Table 1.

TABLE 1
THE NOTATIONS

| parameter | meaning |
|---|---|
| $\mathbb{R}$ | the candidate relaying set without filtering |
| $\mathbb{R}^*$ | the final candidate relaying set after filtering |
| $V_{\mathbb{R}_1^*}$ | the set of nodes in $\mathbb{R}_1^*$. |
| $T$ | waiting time in time-based opportunistic routing |
| $DT_{G(1,2,\ldots,n)}$ | relaying delay of relaying network $G(1,2,\ldots,n)$ |
| $P_{G(1,2,\ldots,n)}$ | packet delivery ratio of relaying network $G(1,2,\ldots,n)$ |
| $P_i$ | the packet delivery ratio of the *i*th priority node in $\mathbb{R}^*$ |
| $\Delta DT_{G(1,2,\ldots,n)}^i$ | the variation of $DT_{G(1,2,\ldots,n)}$ when the packet delivery ratio of *i*th priority node changes |
| $\Delta DT_{G(1,2,\ldots,n)}^{(i,j)}$ | the difference of the $DT_{G(1,2,\ldots,n)}$ variation between any two relaying nodes in $\mathbb{R}^*$ |
| $U$ | the node utility calculated in the first stage of opportunistic routing |
| $ETX_{one\text{-}hop}$ | one-hop ETX for each relaying nodes in $\mathbb{R}^*$ |
| $U_i^*$ | the node utility of the relaying nodes in $\mathbb{R}^*$ when taking the $ETX_{one\text{-}hop}$ into account |
| $neib_i$ | the neighbor matrix of *i*th node in $\mathbb{R}$ |
| $D_{G(1,2,\ldots,n)}$ | the result of (10) of the relaying network $G(1,2,\ldots,n)$ |
| $t_{G(1,2,\ldots,n)}$ | the ETX of the relaying network $G(1,2,\ldots,n)$ |
| $DT_{G(1,2,\ldots,n)}^*$ | network relaying delay when taking $t_{G(1,2,\ldots,n)}$ into account |
| $U_{G(1,2,\ldots,n)}$ | utility of the relaying network $G(1,2,\ldots,n)$ |
| $U_{G(1,2,\ldots,n)}^*$ | utility of the relaying network $G(1,2,\ldots,n)$ when taking $t_{G(1,2,\ldots,n)}$ into account |
| $U_{G(1,2,\ldots,n)}^F$ | final utility of relaying network $G(1,2,\ldots,n)$ |
| $v_{rx}$ | relative variance of parameter *x* |

*B. The calculation model of network relaying delay and packet delivery ratio*

For investigating the performance of the relaying networks, in this section, we will introduce the calculation model of relaying delay and packet delivery ratio of the relaying network. For the time-based coordination scheme, the relaying delay is mainly caused by overhearing the high priority node's ACK message. For better understanding the relaying delay of the time-based coordination scheme, in the following, we introduce the principle of the time-based coordination scheme in detail. The principle can be found in Fig. 3.

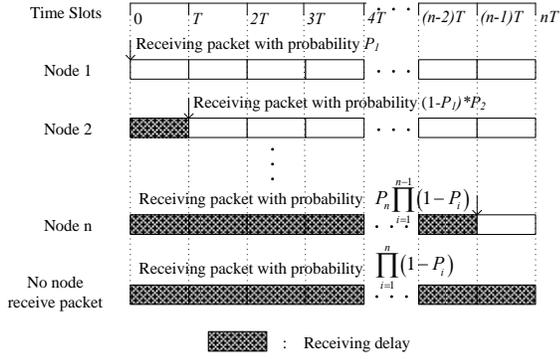

Fig. 3. The principle of the time-based coordination scheme

As shown in Fig. 3, in time-based coordination scheme, the high priority node has high priority to relay data packet to the next hop relaying nodes, the low priority nodes overhear the ACK messages from the high priority nodes. The node priority is determined based on the node utility $U$ which is calculated in the first stage of opportunistic routing algorithm (the different stages of opportunistic routing is introduce in Section I). After the sender sends the data packet to the candidate relaying nodes, the first priority node will check if it receives the data packet. If yes, this node will be the new sender immediately and broadcasts the ACK message to other candidate relaying nodes; the candidate relaying nodes which receive this message will drop the data packet that received from the sender. If the first priority node fails to receive the data packet, then after time $T$ (which is called the waiting time, in [18], this time is set to 45$ms$), the second priority relaying node begins the same process as the first priority node. This process will be repeated until one of the candidate relaying nodes receives the data packet or none of the node receives the data packet. So the average one-hop relaying delay after one transmission try can be calculated as:

$$DT_{G(1,2,\ldots n)} = \left( \sum_{i=1}^{n-1} i P_{i+1} \prod_{j=1}^{i}(1-P_i) + n \prod_{i=1}^{n}(1-P_i) \right) T \quad (1)$$

where $n$ is the degree of the relaying network, $i$ is the priority of each node in the relaying network, $P_i$ is the packet delivery ratio of the $ith$ priority node in $\mathbb{R}^*$ and $0 < P_i < 1$, $T$ is the waiting period. The second term in (1) represents that none of the node receives the data packet transmitted from the sender. Based on the average one-hop relaying delay introduced in (1), we can conclude that to the same relaying network, the network relaying delay will be the smallest when the node priorities are determined based on the packet delivery ratio of node (this can be got easily from (1)). The packet deliver ratio of the relaying network used in this paper is defined as the probability that the data packet sent by the sender can be received by at least one node in $\mathbb{R}^*$. So the packet delivery ratio of the relaying network $G(1,2,\ldots n)$ can be calculated as [17]:

$$P_{G(1,2,\ldots n)} = 1 - \prod_{i=1}^{n}(1-P_i). \quad (2)$$

Note that $P_i$ is different with $P_{G(1,2,\ldots n)}$, since $P_i$ is the probability that the $ith$ priority node in $\mathbb{R}^*$ can receive the data packet from sender, and $P_{G(1,2,\ldots n)}$ is the probability that the data packet sent by the sender can be received by at least one node in $\mathbb{R}^*$. From (1) and (2), we can conclude that even the $s$-networks can be derived from the $o$-networks, the relaying delay and the network packet delivery ratios of these two kinds of networks are different. In the next section, we will investigate the properties of the relaying networks in detail.

## IV. PROPERTIES OF THE RELAYING NETWORK

In this section, based on the calculation model of network relaying delay and network packet delivery ratio that proposed in Section III, we investigate the properties of the relaying network in detail. The properties are divided into in-network properties and inter-network properties. These network properties can be used during the relaying network selection. In the opportunistic routing, for determining the priorities of the candidate relaying nodes, some different performance metrics are used based on different application purposes. These metrics can be divided into two different categories: 1) the packet delivery ratio based metrics, such as the ETX [17], the link correlation [18], etc; and 2) not the packet delivery ration based metrics, such as the distance to the destination nodes, the residual energy, the interference, etc. The network relaying delay of these two different routing algorithms have great difference, since the network relaying delay is affected seriously by the packet delivery ratio of the candidate relaying nodes and their relaying priorities, which will be proved in the following of this section. As shown in (1), the network relaying delay will be different when the node priorities are different to the same network; however, as shown in (2), to the same relaying network, the packet delivery ratio of this relaying network is the same even the node priorities are changed.

### A. Inter-network properties

The inter-network properties represent the properties of the whole relaying network, i.e. the relaying network is regarded as an entirety.

**Corollary 1:** If the $G(V,E)$ is a relaying network, then $E = V(V-1)/2$; otherwise, $E < V(V-1)/2$.

**Proof.** See Appendix A.

For each $\mathbb{R}$, in which the number of candidate relaying nodes is $n$, the number of relaying networks (including the $s$-networks and the $o$-networks) can be calculated as:

$$num = \sum_{i=2}^{n-1} c_n^i. \quad (3)$$

In (3), $c_n^i$ is the number of $i$-degree relaying networks. In this paper, the 1-degree network has been ignored, since the 1-degree network is equal to the deterministic routing.

### B. In-network properties

In the relaying networks, different node parameters, including the packet delivery ratio and the node priority, have different effection on the network performance. For investigating the effection of the node parameter (including the packet delivery ratio and the node priority) on the network performance, in this section, we investigate the in-network properties of the relaying network.

**Definition 5:** To the relaying network $G(1,2,\ldots n)$, the effection of $P_i$ on the network relaying delay is defined as when $P_i$ changes while the packet delivery ratios of the other nodes keep constant, the variation of $DT_{G(1,2,\ldots n)}$, denoted as $\Delta DT^i_{G(1,2,\ldots n)}$.

According to the Definition 5 and (1), the $\Delta DT^i_{G(1,2,...n)}$ (where $1 \leq i \leq n$ and $n$ is the degree of the relaying network) can be calculated as:

$$\Delta DT^i_{G(1,2,...n)} = \left( \sum_{i=1}^{n-1} i P_{i+1} \prod_{j=1}^{i} \left(1 - (P_i + \Delta P)\right) + n \prod_{i=1}^{n} \left(1 - (P_i + \Delta P)\right) \right) T$$

$$= \begin{cases} (1-P_2)\left[ \sum_{j=2}^{n-1} \left( j \cdot P_{j+1} \cdot \prod_{j=3}^{n-1}(1-P_j) \right) \right. \\ \left. + n \cdot \prod_{j=3}^{n}(1-P_j) \right] \Delta P \cdot T, \; i=1 \\ \left[ \left(\prod_{j=1}^{i-1}(1-P_j)\right) \cdot \left[ \sum_{j=i}^{n-1}\left( j \cdot P_{j+1} \cdot \prod_{j=i+1}^{n-1}(1-P_j) \right) \right. \right. \\ \left. \left. + n \cdot \prod_{j=i+1}^{n}(1-P_j) - (i-1) \right] \Delta P \cdot T, \; 1 < i < n \end{cases} \quad (4)$$

where $P_j$ represents the packet delivery ratio of the *jth* relaying node in $G(1,2,…n)$; $n$ is the degree of $G(1,2,…n)$; $\Delta P$ is the variation of the packet delivery ratio $P_i$. Note that the $j$ used in (4) does not the node relaying priority in $\mathbb{R}$, it is the relaying priority in $\mathbb{R}^*$. For instance, if the relaying network is $G(2,6,7)$, then the $P_1$, $P_2$, and $P_3$ in (4) represent $P_2$, $P_6$, and $P_7$, respectively. The coefficient of each term in (4) does not change for the same relaying network. Based on (4), we can calculate the difference of the relaying delay variation between two adjacent relaying nodes $\Delta DT^i_{G(1,2,...n)}$ and $\Delta DT^{i+1}_{G(1,2,...n)}$, which is denoted as $\Delta DT^{(i,i+1)}_{G(1,2,...n)}$. The $\Delta DT^{(i,i+1)}_{G(1,2,...n)}$ can be calculated as follows:

$$\Delta DT^{(i,i+1)}_{G(1,2,...n)} = \Delta DT^i_{G(1,2,...n)} - \Delta DT^{i+1}_{G(1,2,...n)}$$

$$= \left( \prod_{j=1}^{i-1}(1-P_j) \right) \cdot \left[ 1 + (P_i - P_{i+1}) \cdot [1 + (1-P_{i+2}) \right. \quad (5)$$

$$\left. \cdot \left(1 + (1-P_{i+3}) \cdots \left(1 + (1-P_{n-1})(2-P_n)\right) \overset{n-i-2}{\cdots}\right) \right] \Delta P \cdot T$$

Based on (5), we can get the difference of the relaying delay variation between any two relaying nodes, denoted as $\Delta DT^{(i,j)}_{G(1,2,...n)}$, which can be calculated as:

$$\Delta DT^{(i,j)}_{G(1,2,...n)} = \sum_{k=i}^{j-1} \Delta DT^{(k,k+1)}_{G(1,2,...n)}$$

$$= \left( \prod_{j=1}^{i-1}(1-P_j) \right) \cdot \left(1 + (1-P_{i+1}) \right. \quad (6)$$

$$\left. \cdot \left(1 + (1-P_{i+2}) \cdots \left(1 + (P_i - P_j)(2-P_n)\right) \overset{j-i-2}{\cdots}\right) \right) \Delta P \cdot T$$

For instance, for the relaying network $G(1,2,3,7)$, $\Delta DT^{(1,3)}_{G(1,2,3,7)}$ represents the difference of the relaying delay variation between $\Delta DT^1_{G(1,2,3,7)}$ and $\Delta DT^3_{G(1,2,3,7)}$.

**Corollary 3:** To the relaying networks which the priority of the relaying nodes are determined based on the packet delivery ratio based metrics, the higher relaying priorities (i.e. the packet delivery ratio is high), the higher effection on the network relaying delay; i.e. if $i > j$, then $\Delta DT^{(i,j)}_{G(1,2,...n)} > 0$; and if $(i-j) > (i-k)$, then $\Delta DT^{(i,j)}_{G(1,2,...n)} > \Delta DT^{(i,k)}_{G(1,2,...n)}$.

**Proof.** This can be proved directly by (4), (5), and (6).

The Corollary 3 demonstrates that the packet delivery ratios of the high priority relaying nodes have greater effection on the network performance than that of the low priority relaying nodes. Based on (4) and (5), we can derive the Corollary 4 and Corollary 5 as follows.

**Corollary 4:** To the relaying networks which the relaying priorities of the candidate relaying nodes are decided based on the packet delivery ratio based metrics, with the increasing of the network degree, the effection of the same $P_i$ becomes more and more serious, which means if $n > m$, then $\Delta DT^i_{G(1,2,...n)} > \Delta DT^i_{G(1,2,...m)}$ and $\Delta DT^{(i,j)}_{G(1,2,...n)} > \Delta DT^{(i,j)}_{G(1,2,...m)}$.

**Proof.** This can be proved directly by (4), (5), and (6).

For instance, based on Corollary 4, for the relaying networks $G(1,2,3)$ and $G(1,2,3,7)$, the $\Delta DT^{(1,3)}_{G(1,2,3)}$ is smaller than $\Delta DT^{(1,3)}_{G(1,2,3,7)}$ and the $\Delta DT^1_{G(1,2,3)}$ is smaller than $\Delta DT^1_{G(1,2,3,7)}$.

**Corollary 5:** To the relaying network $G(1,2,…n)$ which the priorities of the candidate relaying nodes are decided based on the packet delivery ratio based metrics, with the decreasing of the relaying priority, if $n \to \infty$, then $\Delta DT^{(i,i+1)}_{G(1,2,...n)} \to 0$ and $\Delta DT^i_{G(1,2,...n)} \to 0$.

**Proof.** See Appendix B.

The Corollary 5 demonstrates that the effection of the low priority relaying node on the network performance becomes smaller and smaller when the number of node in the relaying network increases.

For the relaying networks which the node relaying priorities are not decided based on the packet delivery ratio relevant metrics, the properties are the same with that of the relaying networks which the node relaying priorities are decided based on the packet delivery ratio. Before investigating the properties of this kind of relaying network, according to (5) and (6), we propose Corollary 6 first.

**Corollary 6:** To the relaying network $G(1,2,…,n)$ which the relaying priorities of the candidate relaying nodes are not decided based on the packet delivery ratio based metrics, if $P_i < P_j$, then the condition that $\Delta DT^{(i,j)}_{G(1,2,...n)} < 0$ is shown as follows:

$$(P_j - P_i) > \frac{1 + (2 - P_{j-1}) \cdot \prod_{k=i+1}^{j-2}(1-P_k)}{(2 - P_{j+1}) \cdot \prod_{k=i+1}^{j-1}(1-P_k)} = \varphi(i,j) > 1 \quad (7)$$

**Proof.** See Appendix C.

As shown in (8), since $P_i$ and $P_j$ are all smaller than 1, so the $(P_j - P_i)$ is smaller than 1, too. So the (8) will not hold. The conclusion in Corollary 6 means that even $P_i < P_j$, then $\Delta DT^{(i,j)}_{G(1,2,...n)} > 0$. Moreover, the Corollary 6 also illustrates that not only the packet delivery ratio but also the relaying priority can affect the network relaying delay.

Based on Corollary 6, we can conclude that to the relaying networks which the priorities of the candidate relaying nodes are decided based on the packet delivery ratio irrelevant metrics, with the decreasing of the relaying priority, the effection of the node packet delivery ratio on the network relaying delay decreases. This means that in the network which the nodes are prioritized based on the packet delivery ratio irrelevant metrics, we can get the same corollaries as that shown in Corollary 3, Corollary 4, and Corollary 5.

According to the properties of the relaying network, the parameters of node which the relaying priority is high, has greater effection on the transmission delay than that of the node which the priority is low. So for reducing the transmission delay, the high priority relaying nodes should have higher packet delivery ratios than that of the low priority relaying nodes. This conclusion is similar to the conclusions in [17] and [26]. In [26], the authors illustrate that the node's packet delivery ratio which is at the end of the communication link has great effect on the energy consumption; the communication link which this packet delivery ratio is low will deteriorate the routing performance greatly. The authors in [17] use the ETX which relates to all the packet delivery ratios in the communication link to evaluate the effection on the routing performance. In this paper, we prove that the packet delivery ratio of the high priority relaying nodes can affect the transmission delay greatly.

Since for reducing the transmission delay, the high priority relaying node should have higher packet delivery ratio than that of the low priority relaying nodes, however, this is not always hold in the algorithms which the node priority is not determined based on the packet delivery ratio based metrics. In these algorithms, the high relaying priority does not mean small packet delivery ratio. For instance, when the performance metric is residual energy, the node which has large residual energy may not have higher packet delivery ratio than the nodes which have small residual energy. Therefore, for reducing the relaying delay, one approach is re-setting the relaying priority based on the packet delivery ratio. However, this will deteriorate the routing performance, because the node which the residual energy is large may have low relaying priority that determined based on the packet delivery ratio. So to these algorithms, for taking both the node utility that calculated in the first stage of opportunistic routing and the packet delivery ratio into account, the node priority needs to be re-calculated.

Assuming that the utility of *ith* candidate relaying node which calculated in the first stage of the opportunistic routing is $U_i$ ($U_i$ does not take the packet delivery ratio into account), and the packet delivery ratio of this node is $P_i$; according to the definition of ETX in [17], we define the one-hop ETX for each relaying nodes, denoted as $ETX_{one-hop}$, as follows: $ETX_{one-hop} = 1 / P_i$. Therefore, when taking the packet delivery ratio into consideration, the utilities of the candidate relaying nodes that calculated in the first stage of the opportunistic routing will deteriorate; the lower of the packet deliver ratio, the more serious deterioration is. So the new utility which has taken the packet delivery ratio into account can be calculated as:

$$U_i^* = U_i / ETX_{one-hop} = U_i \cdot P_i \quad (8)$$

The (8) demonstrates that when taking the packet delivery ratio into account, the utility of relaying node *i* deduces to $U_i^*$ from $U_i$. The new priorities of the candidate relaying nodes will be determined based on the value of $U_i^*$. An example can be found in Table 2. As shown in Table 2, when taking both the packet delivery ratio and the residual energy into account, node *b* has better performance than node *a* and node *c*. In Table 2, we can find that the high priority node determined by (8) has both high packet delivery ratio and residual energy.

TABLE 2.
AN EXAMPLE

| node | a | b | c | d | e |
|---|---|---|---|---|---|
| residual energy (%) | 0.9 | 0.87 | 0.83 | 0.79 | 0.75 |
| packet delivery ratio (%) | 0.65 | 0.78 | 0.8 | 0.69 | 0.57 |
| priority decided by residual energy | 1 | 2 | 3 | 4 | 5 |
| priority decided by packet delivery ratio | 4 | 2 | 1 | 3 | 5 |
| priority decided by (8) | 3 | 1 | 2 | 4 | 5 |

V. DELAY BASED DUPLICATE TRANSMISSION AVOID (DDA) COORIDNATION SCHEME

In Section III, we introduce the network model and the calculation model of the network relaying delay and network packet delivery ratio; in Section IV, we investigate the properties of the relaying network, including the inter-network properties and in-network properties. In this section, based on the conclusions in Section III and Section IV, we propose the relaying network recognition algorithm (RNR) and delay based duplicate transmission avoid (DDA) coordination scheme.

*A. Relaying network recognition algorithm*

In Section III, we introduce the definition of the relaying network, which is the fully connected sub-network of $G(V_\mathbb{R}, E_\mathbb{R})$. The relaying networks include the *s*-networks and *o*-networks; moreover, the *s*-networks can be derived form the *o*-networks. However, how to judge whether the nodes in $\mathbb{R}^*$ can construct a relaying network or not has not been investigated sufficiently. In this section, based on the conclusion in Corollary 1, we propose a relaying network recognition algorithm (RNR) to estimate whether any *n* nodes can constitute a relaying network or not and distinguish the relaying network is *s*-network or *o*-network.

Before introducing RNR, we first define the neighbor matrix for each candidate relaying node. Assuming that there are *m* nodes in $\mathbb{R}$, for node *i*, the neighbor matrix can be expressed as:

$$\begin{matrix} 1 \ 2 \ 3 \ 4 \ \cdots \ i \ \cdots \ m \\ neib_i = [0 \ 1 \ 0 \ 0 \ \cdots \ 1 \ \cdots \ 1] \end{matrix} \quad (9)$$

In (9), if the node *j* has bi-directional communication link with node *i*, then the *jth* value in $neib_i$ will be "1"; otherwise, this value will be "0". In RNR, we regard that node *i* is a neighbor of itself. For estimating the existence of the relaying network, we define a sum operator between any two neighbor matrixes as follows.

**Definition 7:** For two neighbor matrixes which only contain "0" and "1", the "+" between two neighbor matrixes $neib_i$ and $neib_j$ is defined as:

$$D_{(i,j)} = neib_i + neib_j = \sum_{k=1}^{m} \left( neib_i(k) \wedge neib_j(k) \right). \quad (10)$$

where "$\wedge$" is the "and" operator in Boolean algebra. For instance, to the matrixes [1 0 0 1 1 1] and [0 1 0 1 1 0], based on (10), the summary of these two matrixes will be 2. According to Definition 7, we can estimate whether any *n*-degree network is the relaying network or not.

**Corollary 7:** For any network $G(V_\mathbb{R}, E_\mathbb{R})$ which the network degree is *n*, if $D_{G(V_\mathbb{R}, E_\mathbb{R})} \geq n$, then the network is the relaying network; otherwise, the network is not the relaying network.

**Proof.** See Appendix D.

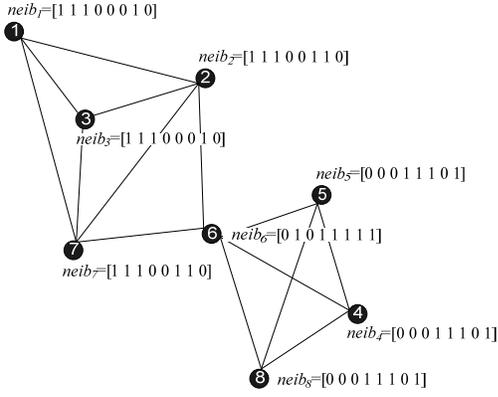

Fig. 4. The neighbor matrixes of the candidate relaying nodes in Fig. 2(a)

For instance, the neighbor matrixes of the candidate relaying nodes in Fig. 2(a) are shown in Fig. 4. As shown in Fig. 4, according to the Definition 7, $D_{(1,2,3)}=4$, which is larger than the network degree of $G(1,2,3)$, so based on the Corollary 7, we can conclude that $G(1,2,3)$ is a relaying network. However, since $D_{(2,5,6)}=1$, which is smaller than the network degree of $G(2,5,6)$, so $G(2,5,6)$ is not a relaying network. The rest of the relaying networks can be gotten by the same process based on the conclusions of Definition 7 and Corollary 7. Note that the relaying networks gotten from Corollary 7 include both the *s*-networks and the *o*-networks. The Corollary 7 is only the algorithm to estimate whether the network is a relaying network or not; it can not distinguish the relaying network is *s*-network or *o*-network. Therefore, we propose the Corollary 8 which can be used to distinguish different kinds of relaying networks.

**Corollary 8:** For any relaying network $G(V_{\mathbb{R}^*})$ which the network degree is $n$, if $D_{G(V_{\mathbb{R}^*})} = n$, then the network $G(V_{\mathbb{R}^*})$ is an *o*-network; otherwise, if $D_{G(V_{\mathbb{R}^*})} = n > m$, where $m$ is the degree of $G(V_{\mathbb{R}^*})$, then $G(V_{\mathbb{R}^*})$ is a *s*-network, and the degree of the *o*-network that $G(V_{\mathbb{R}^*})$ is derived from is $n$; moreover, based on (2), the number of the relevant *m*-degree *s*-network is $c_n^m$.

**Proof.** See Appendix D.

For instance, in Fig. 4, $D_{(1,2,3)}=4$ and the degree of $G(1,2,3)$ is 3, so $G(1,2,3)$ is *s*-network and derived from an *o*-network which the network degree is 4. Additionally, the number of 3-degree relevant *s*-network of $G(1,2,3)$ is 4. In Fig. 4, since $D_{(1,2,3,7)}=4$ which is equal to its network degree, so the network $G(1,2,3,7)$ is an *o*-network.

The relaying network recognition algorithm is shown as follows.

| **Algorithm 1:** The Relaying Network Recognition (RNR) Algorithm |
|---|
| 1. candidate relaying node $i$ calculates the neighbor matrix $neib_i$; |
| 2. if $D_{G(V_{\mathbb{R}^*})_n} = n \rightarrow G(V_{\mathbb{R}^*})_n$ is the *o*-network; |
| 3. if $D_{G(V_{\mathbb{R}^*})_m} = n > m \rightarrow G(V_{\mathbb{R}^*})_m$ is the *s*-network; |
| 4. if $D_{G(V_{\mathbb{R}^*})_n} < n \rightarrow G(V_{\mathbb{R}^*})_n$ is not the relaying network. |

*B. Delay based duplicate transmission avoid (DDA) coordination scheme*

After the recognition of the relaying network, we need to decide which relaying network is the most appropriate one as the final relaying network. The nodes in the selected relaying network will be the final relaying nodes and the other nodes in $\mathbb{R}$ will be deleted.

As talked in Section I, for improving the performance of the opportunistic routing, during the relaying network selection, the following properties of the relaying network should be met as much as possible: 1) the relaying delay of the relaying network should be as small as possible; 2) the packet delivery ratio of the relaying network should be as large as possible; 3) the network in which the node utilities (i.e. the utility is calculated in the first stage of the opportunistic routing) are high should be selected as much as possible for guaranteeing high network performance. Therefore, in the relaying network selection, not only the network packet delivery ratio and the network relaying delay, but also the node utilities in the relaying network should be taken into account.

Based on (1) and (2), the network relaying delay and network packet delivery ratio can be calculated, respectively. Similar to the Expect Transmission Count (ETX) of relaying node which is defined in [17], according to the network packet delivery ratio, we define the one-hop ETX of the relaying network $G(1,2,…,n)$, which can be expressed as:

$$t_{G(1,2,...n)} = \frac{1}{P_{G(1,2,...n)}} = \frac{1}{1-\prod_{i=1}^{n}(1-P_i)} \quad (11)$$

where $P_i$ is the packet delivery ratio of node $i$ in the relaying network. When takes the network ETX into account, the network relaying delay deteriorates, so the network relaying delay which takes the network ETX into account can be calculated as:

$$DT^*_{G(1,2,...,n)} = DT_{G(1,2,...n)} \cdot t_{G(1,2,...n)}$$
$$= \frac{\left(\sum_{i=1}^{n-1} iP_{i+1}\prod_{j=1}^{i}(1-P_j) + n\prod_{i=1}^{n}(1-P_i)\right)T}{1-\prod_{i=1}^{n}(1-P_i)} \quad (12)$$

Similarly to the analysis in Section III, during the relaying network selection, the relaying network which has good performance on both the network relaying delay and the node utilities should have high priority to be selected as final relaying network. For evaluating the effection of the node utilities on the network performance, we define and calculate the network utility $U_{G(1,2,…n)}$ as follows.

For the relaying network $G(1,2,…,n)$, considering the packet delivery ratios and utilities of nodes in the relaying network, which is calculated in the first stage of opportunistic routing, the network utility $U_{G(1,2,…n)}$ varies; this can be expressed in (13):

$$U_{G(1,2,...,n)} = \begin{cases} U_1, & \text{the probability is } P_1 \\ U_2, & \text{the probability is } (1-P_1)P_2 \\ \vdots \\ U_n, & \text{the probability is } \prod_{i=1}^{n-1}(1-P_i)P_n \\ 0, & \text{the probability is } \prod_{i=1}^{n}(1-P_i) \end{cases} \quad (13)$$

where $U_i$ means the utility of *ith* relaying nodes that calculated in the first stage of opportunistic routing (demonstrate in

Section I). Therefore, for a relaying network which the network degree is $n$, the average network utility can be calculated as:

$$\bar{U}_{G(1,2,\ldots,n)} = U_1 \cdot P_1 + \sum_{i=2}^{n}\left(U_i \cdot \prod_{j=1}^{i-1}(1-P_j)P_i\right) \quad (14)$$

The (14) is the average network utility of network $G(1,2,\ldots,n)$ on one transmission try. Similar to the network relaying delay, when taking the network ETX which calculated in (11) into account, this utility deteriorates. According to (12), the network utility which takes the network ETX into account can be calculated as:

$$U^*_{G(1,2,\ldots,n)} = \bar{U}_{G(1,2,\ldots,n)} / t_{G(1,2,\ldots,n)}$$
$$= \left(U_1 \cdot P_1 + \sum_{i=2}^{n}\left(U_i \cdot \prod_{j=1}^{i-1}(1-P_j)P_i\right)\right) \cdot \left(1 - \prod_{i=1}^{n}(1-P_i)\right) \quad (15)$$

Based on (12) and (15), we can find that for each relaying network, two network parameters should be taken into account during the relaying network selection: (1) the network relaying delay $DT^*_{G(1,2,\ldots,n)}$ which takes the network ETX into account and (2) the network utility $U^*_{G(1,2,\ldots,n)}$ which takes the node utility and network ETX into account. The selected relaying network should have high quality performance on both of these two metrics.

In this paper, for achieving this purpose, we introduce the weight based optimal approach into the final network utility calculation, which can be expressed as:

$$U^F_{G(1,2,\ldots,n)} = \omega_{DT} \cdot DT^*_{G(1,2,\ldots,n)} + \omega_U \cdot U^*_{G(1,2,\ldots,n)} \quad (16)$$

where $\omega_{DT}$ is the weight of $DT^*_{G(1,2,\ldots,n)}$, $\omega_U$ is the weight of $U^*_{G(1,2,\ldots,n)}$.

For the weight based algorithm, the first important issue is to determine the weights for each performance metrics. To the metrics of the relaying network, there is a fact that the metric (i.e. $DT^*_{G(1,2,\ldots,n)}$ and $U^*_{G(1,2,\ldots,n)}$) which the variation rate is large has greater effection on the network performance than the metric which the variation rate is small. For instance, as the parameters shown in Table 3, since the values of $U^*_{G(1,2,\ldots,n)}$ between different relaying networks are similar, so which $U^*_{G(1,2,\ldots,n)}$ is chosen as the final relaying network has small effection on the network performance. However, for different relaying networks, the values of $DT^*_{G(1,2,\ldots,n)}$ are quite different, so which $DT^*_{G(1,2,\ldots,n)}$ is chosen has great effection on the network performance. Based on this conclusion, one of the feasible approaches is to use the variances of $DT^*_{G(1,2,\ldots,n)}$ and $U^*_{G(1,2,\ldots,n)}$ as the weights in (16).

However, as shown in [31], if we use the values of $DT^*_{G(1,2,\ldots,n)}$ and $U^*_{G(1,2,\ldots,n)}$ that calculated in (12) and (15), and the variances of $DT^*_{G(1,2,\ldots,n)}$ and $U^*_{G(1,2,\ldots,n)}$ in (16) directly, there may have problems. Because: 1) the final network utility will be mainly decided by the metric which its value is large; for instance, in Table 2, since the value of $DT^*_{G(1,2,\ldots,n)}$ is much larger than $U^*_{G(1,2,\ldots,n)}$, so the value of $U^F_{G(1,2,\ldots,n)}$ will be mainly decided by $DT^*_{G(1,2,\ldots,n)}$; 2) the variance is affected seriously by the value of the metric, so it can not reflect the practical variation rate of the metric; for instance, as shown in Table 3, the variance of $U^*_{G(1,2,\ldots,n)}$ is larger than that of $DT^*_{G(1,2,\ldots,n)}$; however, when taking the values of the metrics into account, the variation rate of $U^*_{G(1,2,\ldots,n)}$ is smaller than that of $DT^*_{G(1,2,\ldots,n)}$ in fact. So when we choose the next hop relaying network, the $DT^*_{G(1,2,\ldots,n)}$ should has greater effection on the routing performance than that of the $U^*_{G(1,2,\ldots,n)}$. This is because variance is the absolute difference between different parameters, so it is affected seriously by the values of the parameters. Therefore, in this paper, for investigating the effection of different metrics on the routing performance, we propose the relative variance ($rv$) which takes the average of the parameter into account and use the relative variance as the weight shown in (16).

TALBE 3.
AN EXAMPLE

| network | a | b | c | variance | rv |
|---|---|---|---|---|---|
| $U^*_{G(1,2,\ldots,n)}$ | 51 | 52 | 53 | 0.67 | 0.00074 |
| $DT^*_{G(1,2,\ldots,n)}$ | 0.27 | 0.68 | 0.49 | 0.028 | 0.366 |

The relative variance is defined as:

$$v_{rx} = \frac{1}{n}\sum_{i=1}^{n}\left(\frac{x_i - \bar{x}}{\bar{x}}\right)^2 \quad (17)$$

where $x$ represents $DT^*_{G(1,2,\ldots,n)}$ and $U^*_{G(1,2,\ldots,n)}$, $\bar{x}$ is the average of $x$, $n$ is the number of relaying networks. In the relative variance, the value of (17) can reflect the effection of different parameters on the routing performance accurately. This can be found in Table 3. In Table 3, even the variance of $U^*_{G(1,2,\ldots,n)}$ is larger than that of $DT^*_{G(1,2,\ldots,n)}$, the relative variance of $DT^*_{G(1,2,\ldots,n)}$ is larger than that of $U^*_{G(1,2,\ldots,n)}$, which is consist with the effection of the metric on the routing performance.

For evaluating the difference between the relative variances of these two metrics, we define the parameter resolution ratio $\xi$ as:

$$\xi = \begin{cases} v_{rDT} / v_{rU}, & v_{rDT} > v_{rU} \\ 1, & v_{rDT} = v_{rU} \\ v_{rU} / v_{rDT}, & v_{rDT} < v_{rU} \end{cases} \quad (18)$$

From (18), we can find that $\xi \geq 1$, the larger $\xi$ is, the larger difference between the relative variances of these two parameters. For the network utility calculated in (16), with the increasing of $\xi$, the effection of the metric which the relative variance is large on the network utility increases, and the effection of the parameter which the variance is small decreases. When the $\xi$ is small, the effection of these two parameters on the network utility is similar.

For the first issue, if we use the values of metrics directly in the network utility calculation, then there will have problems. For instance, as the metrics shown in Table 3, since the relative variance of Metric_1 is smaller than that of the Metric_2, so according to the analysis above, the network utility should be affected mainly by the Metric_2; however, the fact is that the network utilities are decided mainly by Metric_1, i.e. the network which the value of Metric_1 is the largest will have the highest network utility. According to the network utility defined in (16), the priorities of the network utilities are: *network_c→network_b→network_a*, which is the same as the priorities of Metric_1. This is not consistent with the analysis above. The reason is that the value of Metric_1 is much larger than that of the Metric_2. When the difference between Metric_1 and Metric_2 is too large, it will cover up the effection of Metric_2 on the relaying network selection. For solving this issue, in [31], the authors map the different order of magnitudes parameters to the same order of magnitude; in this paper, considering the fact that for each performance metric, there has an order number relates to each relaying network, so we introduce the order number of the relaying network into the network utility calculation. For instance, as the values of Metric_2 shown in Table 4, the order numbers of the relaying networks relate to Metric_2 are: network_a→1, network_b→3, and network_c→2, respectively. The large order number means that the related metric's value is large in the relaying network, vice versa. So in this paper, the value of parameter shown in (16) will be replaced by the order number of each relaying network, which can be expressed as:

$$U^F_{G(1,2,\ldots,n)} = v_{rDT} \cdot n^i_{DT_{G(1,2,\ldots,n)}} + v_{rU} \cdot n^i_{U_{G(1,2,\ldots,n)}} \quad (19)$$

where $n^i_{DT_{G(1,2,\ldots,n)}}$ is the order number of $G(1,2,\ldots,n)$ relates to $DT$, $n^i_{U_{G(1,2,\ldots,n)}}$ is the order number of $G(1,2,\ldots,n)$ relates to $U$. The network utility will be decided by (19), which can be found in Table 4. In Table 4, the network utility of network_b is larger than that of network_c, which is consistent with the analysis above. In Table 4, we also present the network utilities that calculated based on the algorithm proposed in [31] which is the weight based algorithm and [34] which is the fuzzy logic based algorithm. From Table 4, we can find that the priorities of the relaying networks that calculated by (19) are the same as that calculated by [31] and [34].

TABLE 4.
AN EXAMPLE

| network | a | b | c | rv |
|---|---|---|---|---|
| Metric_1 | 29 | 45 | 63 | 0.0925 |
| Order number of Metric_1 | 1 | 2 | 3 | |
| Metric_2 | 0.27 | 0.68 | 0.49 | 0.122 |
| Order number of Metric_2 | 1 | 3 | 2 | |
| Utility calculated by (16) | 2.72 | 4.25 | 5.89 | |
| Utility calculated by (19) | 0.3365 | 0.551 | 0.3995 | |
| Utility calculated by [31] | 0.06 | 0.125 | 0.118 | |
| Utility calculated by [34] | 0.448 | 0.529 | 0.517 | |

Based on (18) and (19), we can derive the property of this algorithm as follows. The network utility calculated by (19) relates to both the weight of the metric and the priority of the relaying network. Assuming that there are two relaying networks, for the network_1, the order number based on $DT$ is $n_i$ and the order number based on $U$ is $n_j$; for the network_2, the order numbers relate to these two Metrics are $n_m$ and $n_k$, respectively. Let $\Delta^n_{DT} = |n_i - n_j|$, $\Delta^n_U = |n_m - n_k|$, $\xi = \alpha$, and $v_{rDT} > v_{rU}$, then we can derive the property of this algorithm as follows.

**Corollary 9.** If $\Delta^n_U / \Delta^n_{DT} < \alpha$, the utility will be decided mainly by $DT$, and if $\Delta^n_U / \Delta^n_{DT} > \alpha$, then the utility will be decided mainly by $U$; vice versa.

**Proof.** See Appendix E.

An example can be found in Fig. 6. The values of the metrics in Fig. 6 are the same as that shown in Table 3. As shown in Fig. 6(a), for the network_2 and network_3, since $\Delta_{DT} = 1$ and $\Delta_U = 1$, so $\Delta^n_U / \Delta^n_{DT} = 1$; since $\xi = 1.32 > \Delta^n_U / \Delta^n_{DT}$, so in Fig. 6(a), the network utility will be decided mainly by the value of $DT$. Therefore, in Fig. 6(a), the relaying priority of network_2 is 1 and the priority of network_3 is 2, which is the same as the order of $DT$. However, as shown in Fig. 6(b), for network_2 and network_3, since $\Delta^n_U / \Delta^n_{DT} = 2$ which is larger than $\xi$, so the network utility will be decided mainly by the value of $U$. Therefore, in Fig. 6(b), the relaying priorities of network_2 and network_3 are 2 and 1, respectively; this is the same as the order of $U$. The relaying priorities of network_2 and network_3 are opposite in these two figures.

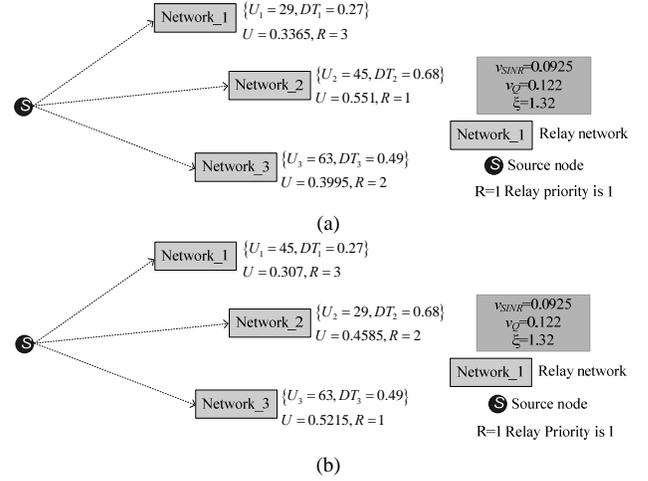

Fig. 6. An example of the relaying network prioritize and selection algorithm

When the relaying network is selected by the algorithm introduced above, the nodes in the relaying network will relay the data packet based on the relaying priority that calculated in Section IV. The relaying is time-based, which has been introduced in [17], [18], and [22] in detailed; the waiting timer is set to 45ms, which is the same as that shown in [18]. When the node in the relaying network relays the data packet to the next hop relaying network, the processes are the same as that shown above until the data packet is received by the destination node. The process of the DDA can be found below.

**Algorithm 2:** DDA coordination scheme

1. each relaying network calculated the network ETX based on (11);
2. based on (12) and (15), the network relaying delay $DT^*_{G(1,2,\ldots,n)}$ and network utility $U^*_{G(1,2,\ldots,n)}$ are calculated;
3. the source node calculate the variances of network relaying delay and network utility, i.e. $\xi_{DT^*_{G(1,2,\ldots,n)}}$ and $\xi_{U^*_{G(1,2,\ldots,n)}}$, respectively;
4. applying the Corollary 2, Corollary 3, and Corollary 4 to pre-select the relaying network;
5. based on (19), the final network utility $U^F_{G(1,2,\ldots,n)}$ is calculated;
6. the relaying network which has highest final network utility will be chosen as the relaying network.

## VI. SIMULATION AND DISCUSSION

In this section, we will evaluate the performance of DDA coordination scheme. We compare the performance of DDA with ExOR [17] and SOAR [18], respectively. The variation parameters are the number of nodes and the number of CBR connections. The number of CBR connections represents the traffic load of the network. The parameters of the simulation environments are shown in Table 5.

TALBE 5.
SIMULATION PARAMETERS

| simulation parameter | value |
| --- | --- |
| simulation area | 2000$m$×2000$m$ |
| number of vehicles | 100, 150,…, 300 |
| transmission range | 250$m$ |
| channel data rate | 1Mbps |
| the traffic type | Constant Bit Rate (CBR) |
| number of CBR connections | 20, 40,…, 100 |
| packet size | 512bytes |
| beacon interval | 1$s$ |
| maximum packet queue length | 50 packets |
| MAC layer | IEEE 802.ll DCF |
| simulation tool | NS2 |

The algorithms used in this simulation are ExOR, SOAR, and DDA. The introduction of ExOR and SOAR can be found in [17] and [18], respectively. The DDA is the algorithms that proposed in this paper, the detail of DDA can be found in Section IV and Section V.

The performance matrixes used in this paper are the transmission delay, the packet delivery ratio between sender and the candidate set, and the network throughput: (1) *End-to-End Packet delivery ratio*: the packet delivery ratio is defined as the ratio of the number of packets received successfully by the destination node to the number of packets generated by the source node [26][32]; (2) *End-to-End delay*: the transmission delay of the data packet from the source node to the destination node; (3) *Network throughput*: the network throughput is the ratio of the total number of packets received successfully by the destination node to the number of packets sent by all the nodes during the simulation time [33].

*A. Performance under different network density*

In this section, we evaluate the performance of DDA, SOAR, and ExOR under different network density, i.e. the number of nodes in the network varies. In this simulation, the network load is constant, which means that the number of the CBR connections is set to 60. The results can be found in Fig. 7, Fig. 8, and Fig. 9.

In Fig. 7, the average end to end delays of these three algorithms are presented. In these three algorithms, with the increasing of the number of nodes in the network, the average end to end delay decreases both in these three algorithms. The fewer nodes in the network, the larger decrease is. For instance, in DDA, when the number of nodes in the network increases from 100 to 150, the delay decreases from 780$ms$ to 602$ms$; however, when the number of nodes increases from 250 to 300, the delay decreases from 520$ms$ to 500$ms$. The similar conclusion can be found in SOAR and ExOR. This can be explained as: when the number of node increases, the probability of network portion decreases, so the delay will decrease when the network density increases; when the network density is large enough, then the probability of network portion is quite low, so the decreasing of the transmission delay is slow. Moreover, for the same network density, the end to end delay of DDA is much smaller than that of the other two algorithms. This is because in DDA, the relay nodes are fully connected and the relay network which the delay is the small has high priority to be chosen, so the end to end delay in DDA is the smallest in these three algorithms.

In Fig. 8, the packet delivery ratios of these three algorithms are illustrated. With the increasing of the network density, the packet delivery ratios of these three algorithms increase; the packet delivery ratio of DDA is the largest in these three algorithms. Since with the increasing of the network density, for the sender, more relaying nodes can be found in its transmission range, so according to the calculation in (2), the network packet delivery ratio will increase. Since in DDA, the packet delivery ratio is taken into account during the relaying network selection, so the packet delivery ratio of DDA is the largest. In Fig. 8, when the network density is large enough, this increasing becomes slowly; this is due to when the network density is large enough, the number of candidate relaying nodes is large, so there always exits at least one node can receive the data packet and send it to the destination node, which makes the increasing slow.

The network throughputs of these three algorithms are presented in Fig. 9. From Fig. 9, we can conclude that when the network density increases, the network throughput keeps constant approximately; these values fluctuate in a very small range. For instance, the variation range of DDA is 0.03 approximately and is about 0.02 in SOAR. On one hand, when the network density is small, the packet delivery ratio is small which can be found in Fig. 8, then the probability of retransmission is high; however, the number of hops to the destination is small when the network density is small, which contributes to the number of control packet reduction. On the other hand, when the network density is large, the packet delivery ratio increases; however, the average number of hops to the destination node increases, which causes the number of control packets increasing. So the network throughput keeps stable in these algorithms; moreover, the network throughput of DDA is the best in these three algorithms.

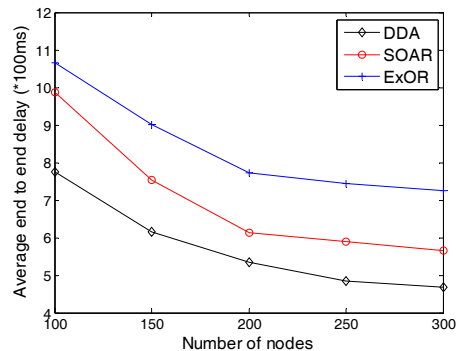

Fig. 7. The average end to end delay under different network densities.

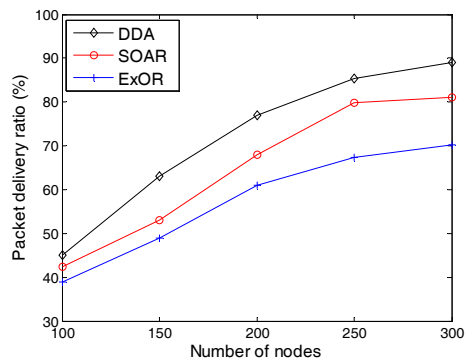

Fig. 8. The packet delivery ratio under different network densities.

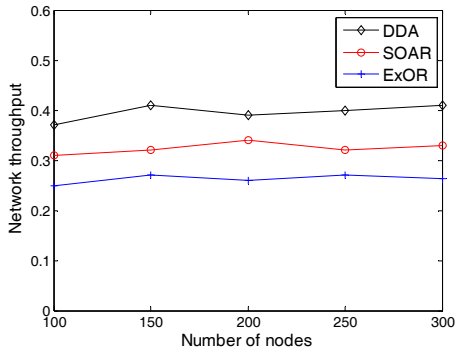

Fig. 9. The network throughput under different network densities.

## B. Performance under different traffic load

In this section, the performance of these three algorithms under different number of CBR connections is presented. In this simulation, the number of nodes in the network is 200 and the number of CBR connections varies. The results can be found in Fig. 10, Fig. 11, and Fig. 12.

In Fig. 10, the average end to end delays of these three algorithms are shown. The results of the end to end delay under different traffic load are different with that of under different network densities. With the increasing of the number of CBR connections, the end to end delay is the smallest when the number of CBR connections is 100; the delay decreases when the number of CBR connections smaller than that and increases when the number of CBR connections larger than that. This is because with the traffic load increasing, when the number of CBR connections is not large enough (for instance, smaller than 100), the network resources are far from saturated, so when the traffic load increases, the end to end delay decreases; however, when the number of CBR connection is large enough, the network becomes saturated or over-saturated, so the network contend becomes more and more serious, which will deteriorate the performance of the algorithms.

When the traffic load increases, the packet delivery ratios of these three algorithms decrease, which can be found in Fig. 11. The reason of the packet delivery ratios decreasing is because when the number of CBR connections increases, the network contend becomes more and more serious. Moreover, similar to the Fig. 10, when the number of CBR connections is small, this decreasing is slow; however, when the number of CBR connections is large, this decreasing is fast. This is because when the number of CBR connections is small, the network resources, such as the buffer of each node, are not saturated, so even the network contend and the network interference increase, the decreasing of the packet delivery ratio is slow. However, when the network resource is saturated or over-saturated, the network interference and the network contend increase, so the decreasing of the packet delivery ratio becomes more and more serious.

The network throughput of these three algorithms under different number of CBR connections is shown in Fig. 12. Different with that shown in Fig. 9, the network throughput shown in Fig. 12 decreases when the network traffic load increases; however, the decreasing of these three algorithms is slight. The decreasing of the network throughput can be explained by Fig. 10 and Fig. 11, when the number of CBR connections increases, on one hand, the end to end delay decreases at first and increases after the inflection point (i.e. 100); on the other hand, when the traffic load increases, the packet delivery ratio decreases; additionally, when the traffic load increases, the network interference, the network contend, and the channel occupation ratio increase seriously, so the network throughput decreases. However, as that shown in Fig. 12, the decreasing of ExOR and SOAR is much faster than that of DDA; moreover, the network throughput of DDA is the largest in these three algorithms. This is because the duplication transmission in the time-based coordination scheme is reduced as much as possible in DDA, which contributes to the increasing of the network throughput.

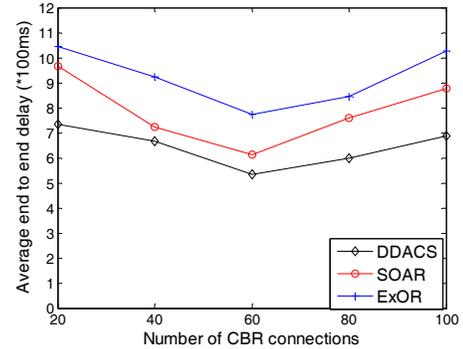

Fig.10. The average end to end delay under different traffic loads.

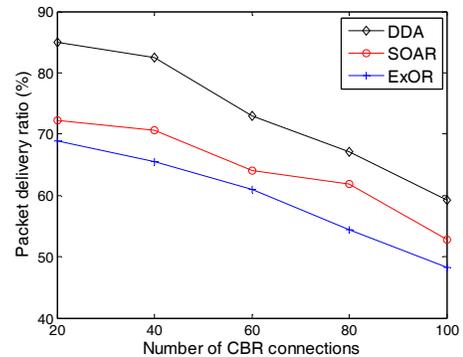

Fig.11. The packet delivery ratio under different traffic loads.

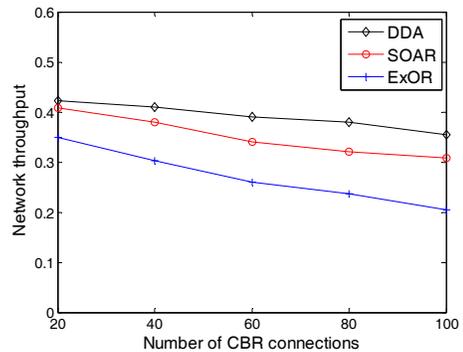

Fig.12. The network throughput under different traffic loads.

## VII. CONCLUSION

In this paper, for reducing the transmission delay and the duplication transmission in the opportunistic routing, we propose the delay based duplication transmission avoid (DDA) coordination scheme for the opportunistic routing. In this coordination scheme, the candidate relaying nodes are divided into different fully connected relaying networks, so the duplication transmission is avoided. Moreover, in this paper, we also propose RNR algorithm which can be used to judge whether the sub-network is fully connected or not. When the fully connected relaying networks are got, then these relaying networks will be used as the basic units in the next hop relaying network selection. In this paper, we prove that the packet delivery ratio of the high priority relaying nodes in the relaying network has greater effection on the relaying delay than that of the low priority relaying nodes. According to this conclusion, in DDA, the relaying networks which the packet delivery ratios of the high priority relaying nodes are high have higher priority than that of the low one. During the next hop relaying network selection, the transmission delay, the network utility, and the packet delivery ratio are taken into consideration. By these innovations, the DDA can improve the network performance greatly than ExOR and SOAR. Moreover, in this paper, the properties of the relaying networks are investigated in detail.